\documentclass{pasj01}

\usepackage{CJKutf8}
\DeclareRobustCommand{\adsout}{\bgroup\markoverwith{\textcolor{blue}{\rule[.5ex]{2pt}{0.4pt}}}\ULon}
\usepackage{ulem,appendix}

\Received{$\langle$reception date$\rangle$}
\Accepted{$\langle$acception date$\rangle$}
\Published{$\langle$publication date$\rangle$}

\begin{document}

\title{ Nuclear Burning in Accretion Flow of Helium-rich matter onto Compact Objects}
\author{Toshikazu \textsc{Shigeyama}\altaffilmark{1}}%
\altaffiltext{1}{Research Center for the Early Universe, Graduate School of Science,  University of Tokyo,
 7-3-1 Hongo, Bunkyo-ku, Tokyo 113-0033, Japan }
\email{shigeyama@resceu.s.u-tokyo.ac.jp}
\author{Akira \textsc{Dohi}\altaffilmark{2,3}}%
\altaffiltext{2}{Astrophysical Big Bang Laboratory (ABBL), Cluster for Pioneering Research, RIKEN, Wako, Saitama 351-0198, Japan }
\altaffiltext{3}{Interdisciplinary Theoretical and Mathematical Sciences Program (iTHEMS), RIKEN, Wako, Saitama 351-0198, Japan }

\KeyWords{accretion --- supernovae: general --- nuclear reactions --- binaries: close --- hydrodynamics}

\maketitle

\begin{abstract}
We investigate the impacts of nuclear burning on the spherically symmetric stationary accretion flow of helium-rich matter on to compact objects. We have already shown the existence of the critical accretion rates for the accretion of CO-rich matter above which the flow truncates in the supersonic region due to nuclear burning in the previous paper \citep{2022ApJ...933...29N}. Here, we show that there are also critical accretion rates for helium-rich matter. While we used empirical formulae for the energy generation rates for carbon burning and oxygen burning without solving the nuclear reaction network in our previous work, we solve a simple nuclear reaction network consisting of 13 elements from $^4$He to $^{56}$Ni to investigate influence of the energy generation from not only triple-$\alpha$ reactions but also the subsequent reactions of synthesized elements. We have also qualitatively confirmed the previous results for CO-rich matter accretion using the revised code with the nuclear reaction network and reported some new findings. 
\end{abstract}

\section{Introduction}
Accretion onto a compact object is an important phenomenon in the evolution of a close binary system. There are many branches of the evolution of such close binary systems\citep{2014LRR....17....3P}. 
 Some of them end up with merging events, i.e., the engulfment of the compact object by the companion star. Eventually, the compact object resides in the core of the companion star, which is composed of He or C and O if the companion star is massive and has evolved into a supergiant. Here, we focus on the accretion flow associated with such a merging system. 
If the companion is a main-sequence star and the compact object is a neutron star, the engulfment results in a quasi-hydrostatic configuration with a degenerate neutron core investigated by \citet{1975ApJ...199L..19T}. The dynamical phase must, however, precede it before the system evolves to the quasi-hydrostatic equilibrium state. If the companion is an evolved star with a core composed of He or C and O, the accretion by a compact object residing in the core after the engulfment will be affected by nuclear burning since the timescales of carbon and/or oxygen burning can be as short as the dynamical timescale and the generated energy can be comparable to the gravitational energy.
In this context, \citet{2022ApJ...933...29N} investigated the impacts of carbon burning and oxygen burning on the accretion flow of CO-rich matter onto compact objects by constructing a series of spherically symmetric stationary accretion models. They found that there are two types of solutions: one is obtained when the accretion rate is small, in which the transonic flow reaches the center, while in the other type of solution for large accretion rates, the flow truncates in the supersonic region at the point where the flow velocity becomes equal to the sound speed again. 
They derived the critical accretion rates that separate the types as a function of the specific enthalpy in the ambient medium for a given mass of the compact object. The actual value of the critical accretion rate onto a compact object with a mass of $M$ is about $10^{33}$ g s$^{-1}$ and decreases with the specific enthalpy and increases with $M$.   

Here we take the same approach as our previous study \citep{2022ApJ...933...29N} but with a small nuclear reaction network which contains 13 species (\atom{He}{}{4}, \atom{C}{}{12}, \atom{O}{}{16}, \atom{Ne}{}{20}, \atom{Mg}{}{24}, \atom{Si}{}{28}, \atom{S}{}{32}, \atom{Ar}{}{36}, \atom{Ca}{}{40}, \atom{Ti}{}{44}, \atom{Cr}{}{48}, \atom{Fe}{}{52}, and \atom{Ni}{}{56}). The inclusion of this reaction network allows us to investigate the accretion of He-rich matter because it is expected that the flow experiences not only triple-$\alpha$ reactions but also the subsequent reactions with heavier elements.

The structure of the paper is as follows. Section \ref{sec:method} presents the governing equations for the spherically-symmetric stationary accretion flow with nuclear burning and briefly describes a method to solve them. Section \ref{sec:results} presents results of numerical calculations for accretion flows with different compositions and investigates the nature of each result in some detail. Section \ref{sec:conclusion} concludes the paper.

\section{Methods}\label{sec:method}
\subsection{Steady transonic solutions}
The governing equations (the conservations of mass, momentum, and energy) are basically the same as those in \citet{2022ApJ...933...29N}, that is,
\begin{eqnarray}
&\dot M=4\pi r^2 v\rho,& \label{eq:rate}\\
&\rho v\frac{dv}{dr}+\frac{dp}{dr}+\frac{GM\rho}{r^2}=0,&\\
&v\left(\frac{d\epsilon}{dr}-\frac{p}{\rho^2}\frac{d\rho}{dr}\right)=\varepsilon_{\rm nuc}(\rho,T)-\varepsilon_\nu(\rho, T).&\label{eq:energy}
\end{eqnarray}

Here $r$ denotes the radial coordinates, i.e., the distance from the center of the compact object (a neutron star or a black hole) with a mass of $M$, $v$ the radial velocity, $\rho$ the mass density, $p$ the pressure, $\epsilon$ the specific thermal energy density,  $\varepsilon_{\rm nuc}(\rho,T)$ the energy generation rate due to nuclear reactions as a function of the density and the temperature $T$,  and $G$ is the gravitational constant. Note that the accretion rate $\dot M$ and the velocity $v$ have negative values since the outward direction is positive. 
We neglect the radius of the compact object. To obtain thermodynamic quantities such as $\epsilon$ and $p$, we utilize the equation of state given by \citet{2000ApJS..126..501T}. We solve these equations for given accretion rate $\dot M$, mass $M$ of the compact object, and the specific enthalpy $h_\infty$ in the ambient matter by integrating the differential equations from the sonic point as was done in \citet{2022ApJ...933...29N}. In this work, the energy loss term $\varepsilon_\nu(\rho, T)$ due to neutrino emission \citep{1996ApJS..102..411I} is included, and we confirm that this term is always negligible in the obtained solutions. The energy generation term is calculated from the results of the nuclear reaction network calculations described in the next subsection.

To solve the above governing equations, we search the radius of the sonic point for a given set of $M$, $\dot M$, and $h_\infty$ assuming no nuclear reactions in the subsonic outer region and obtain the physical values and their derivatives with respect to $r$ at the point. Then, we numerically integrate the governing equations toward the center, including the nuclear reaction network from a point slightly shifted from the sonic point using the physical values and the derivatives. When we integrate outward, we assume that nuclear reactions do not take place. This is found to be a good assumption when $h_\infty$ is low. We will explain the limit in each specific case below.


\subsection{Nuclear reaction network}
The nuclear reaction network implemented in our calculation solves the following rate equations;
\begin{eqnarray}
    \frac{dy_i}{dt}&=&\sum_{j,k=1}^{13}\mathcal{N}^i_{jk}\rho N_{\rm A}\left<\sigma v\right>^i_{jk}y_jy_k+\sum_{j=1}^{13}\mathcal{N}^i_j\left<\sigma v\right>^i_jy_j\\
    &&+\sum_{j,k,l=1}^{13}\mathcal{N}^i_{jkl}\left(\rho N_{\rm A}\right)^2\left<\sigma v\right>^i_{jkl}y_jy_ky_l, 
\end{eqnarray}
where $y_i$ denotes the number fraction of a nuclide $i$ per nucleon, $N_{\rm A}$ the Avogadro number, $\left<\sigma v\right>$ the average reaction rates, which are given by empirical fitting formulae with temperature \citep{Cyburt_2010}, $\mathcal{N}$ are numbers that specify the relevance of the reaction to the nuclide $i$, and $d/dt$ denotes the Lagrange time derivative, which is replaced by $v\partial/\partial r$ in the steady state solutions. The enhancement of reaction rates due to Coulomb correlations is taken into account by using the formulation of \citet{1984PhRvA..29.2033I}. However, the effects of this screening are found to be always negligible in the solutions obtained in this paper due to their low densities. The energy generation rate $\epsilon_{\rm nuc}$ is calculated from the equation
\begin{equation}
    \epsilon_{\rm nuc}=-\sum_{i=1}^{13}\Delta m_ic^2\frac{dy_i}{dt},
\end{equation}
where $\Delta m_i$ is the mass excess of nuclide $i$ and $c$ is the speed of light.

We use the numerical code of \citet{1986A&A...162..103M} to integrate these equations. We have included effects of $\left(\alpha,{\rm p}\right)$ and the reverse reactions \citep{1992ApJ...386L..13S} to the original code. Thus the reactions included in this study become $(\alpha,\gamma),(\alpha,p),(p,\gamma)$, 3$\alpha$, C fusion, O fusion, and their reverse reactions.


\section{Results}\label{sec:results}
\subsection{Accretion of pure He matter}\label{sec:pHe}
First, we investigate the accretions of matter composed of pure He to understand which reactions are influential to the dynamical behavior of the flow, though this composition is unrealistic in the situation that no elements heavier than ${}^4{\rm He}$ are included. 
\subsubsection{Features of flows}\label{sec:fof}
\begin{figure}
 \begin{center}
  \includegraphics[width=80 mm]{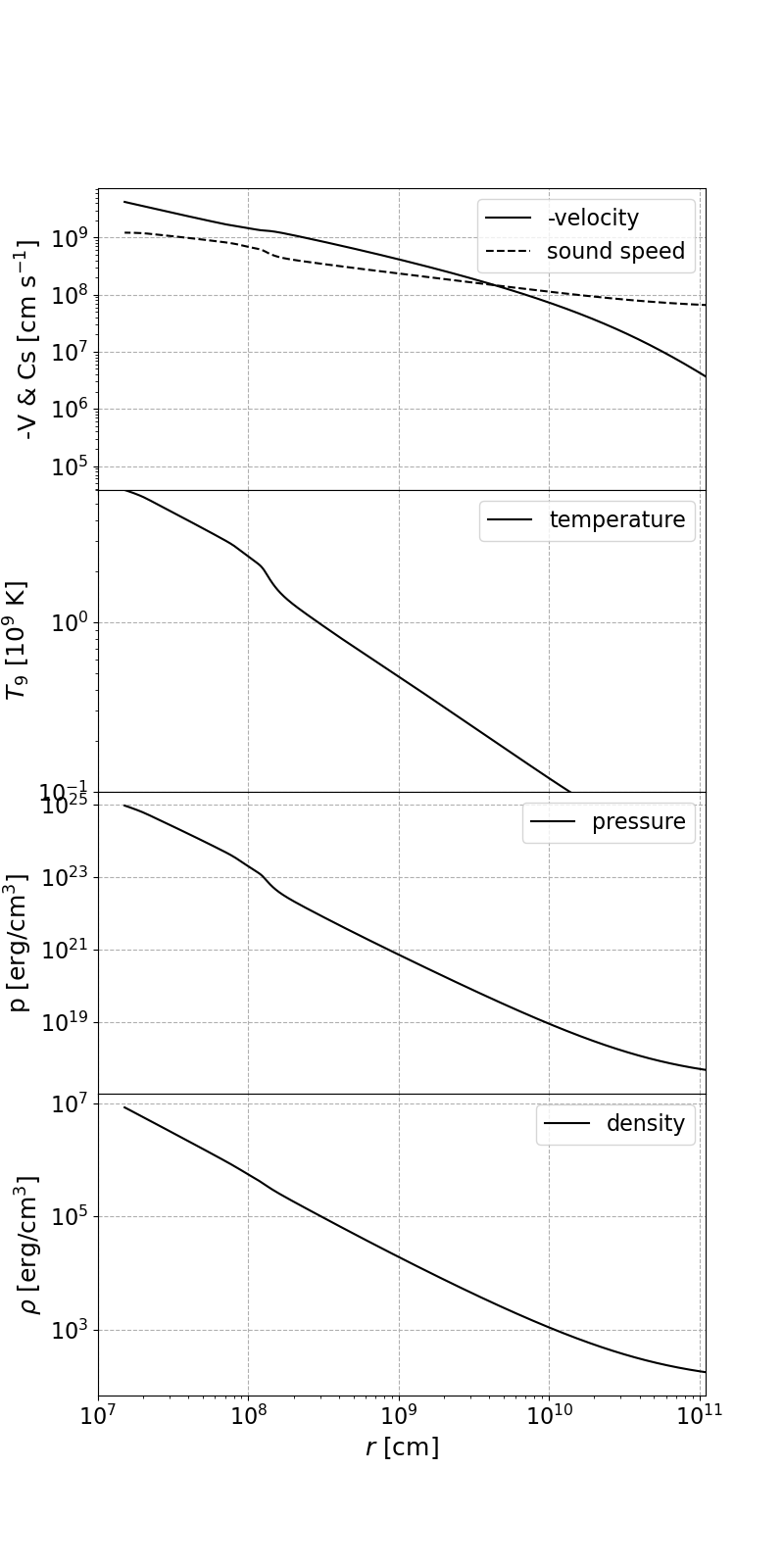}
 \end{center}
 \caption{The distributions of the velocity (with the sound speed), the temperature, the pressure, and the density as functions of radius (from the top to the bottom panels, respectively). The parameters of this model are $M=1.4\ M_\odot$, $|\dot M|=10^{32}$ g s$^{-1}$, and $h_\infty=5.19\times10^{15}$ erg g$^{-1}$. Note that the numerical integration is stopped when the temperature exceeds $6\times10^9$ K.}\label{f:pHe@sar}
\end{figure}

\begin{figure}
 \begin{center}
  \includegraphics[width=80 mm]{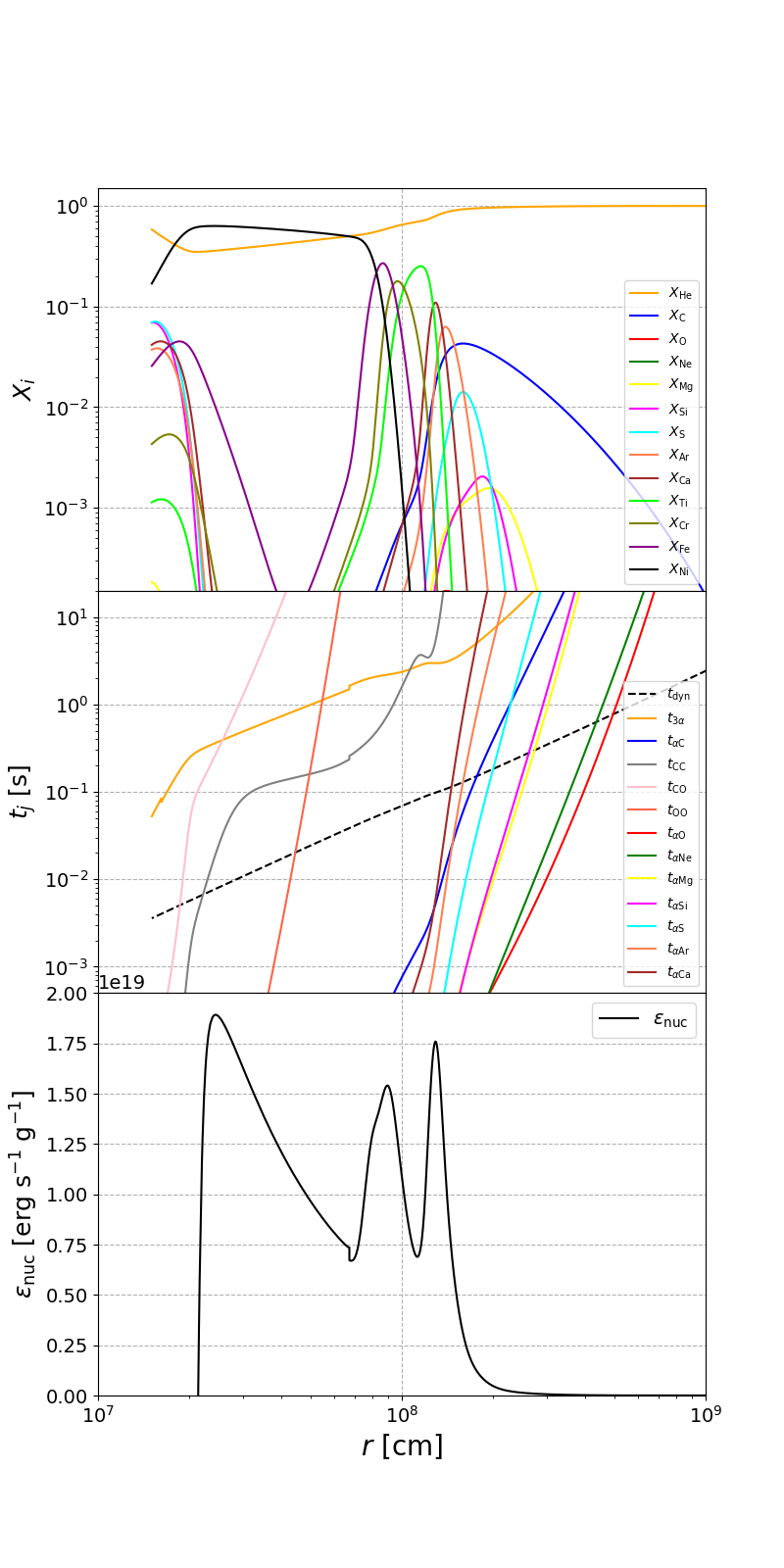}
 \end{center}
 \caption{The distributions of the mass fraction of the 13 elements (the top panel), the time scales of $\alpha$-capture reactions and some other reactions in addition to that of the flow (the middle panel), and the energy generation rate per unit mass from the nuclear reactions (the bottom panel) as functions of radius in the supersonic region. The parameters of this model are the same as in Figure \ref{f:pHe@sar}. The dynamical timescale defined as $r/(-v)$ is shown by dashed line in the middle panel.}\label{f:pHe@sarN}
\end{figure}

When the accretion rate is smaller than the critical one ($\dot M_{\rm cr}(h_\infty, M)$), which is a function of the specific enthalpy $h_\infty$ in the ambient medium and the mass $M$ of the compact object, the flow approaches the center at supersonic speeds. One example is shown in Figures \ref{f:pHe@sar} and \ref{f:pHe@sarN} for a parameter set of $M=1.4\ M_\odot$, $\dot M=10^{32}$ g s$^{-1}$, and $h_\infty=5.19\times10^{15}$ erg g$^{-1}$.  A transonic flow reaching the center is obtained, in which the sonic point is located at $r\sim3.3\times10^9$ cm. Since the temperature in the subsonic region is less than $3\times10^8$ K, no significant nuclear reactions proceed there. The timescale of the triple-$\alpha$ reactions (defined as $t_{3\alpha}=6/((\rho N_{\rm A})^2\left<\sigma v\right>_{3\alpha}y_{\rm He}^2)$) is much longer than the dynamical timescale $t_{\rm dyn}\equiv r/(-v)$ even in the supersonic region near the sonic point as shown in the middle panel of Figure \ref{f:pHe@sarN}. As a result, the peak abundance of C is only $\sim0.07$ in mass fraction (the top panel of the same figure). The abundance of C starts decreasing toward the center where the reaction timescale of $\alpha$ capture of C (defined as $t_{\alpha\rm C}=1/(\rho N_{\rm A}\left<\sigma v\right>_{\alpha\rm C}y_{\rm He})$) becomes shorter than $t_{\rm dyn}$. The abundance of each heavy element has its peak where the destruction timescale of the element by $\alpha$ capture becomes as short as not only the timescale of the production reactions but also that of the accretion flow. A \atom{Ni}{}{56}-rich region appears where the temperature exceeds $4\times10^9$ K and $r<7\times10^8$ cm. Photo-disintegrations of \atom{Ni}{}{56} in the innermost region produce \atom{He}{}{4} and intermediate-mass elements (the top panel of Fig. \ref{f:pHe@sarN}). Note that we stop the numerical integration when the temperature exceeds $6\times10^9$ K above which the nuclear reaction network may cause some numerical problems especially in calculation of the energy generation rate. This treatment does not affect the upstream because the region is supersonic. In the supersonic region, the flow is affected by nuclear energy generation due to C+C fusion reactions and $\alpha$-capture reactions with heavier elements produced by C+C at around $r\sim10^8$ cm where the temperature approaches to $2\times10^9$ K and these reactions start to dominate over the triple-$\alpha$ reactions. The pressure increase toward the downstream is enhanced to decelerate the flow there as shown in Figure \ref{f:pHe@sar}. 

\begin{figure}
 \begin{center}
  \includegraphics[width=80 mm]{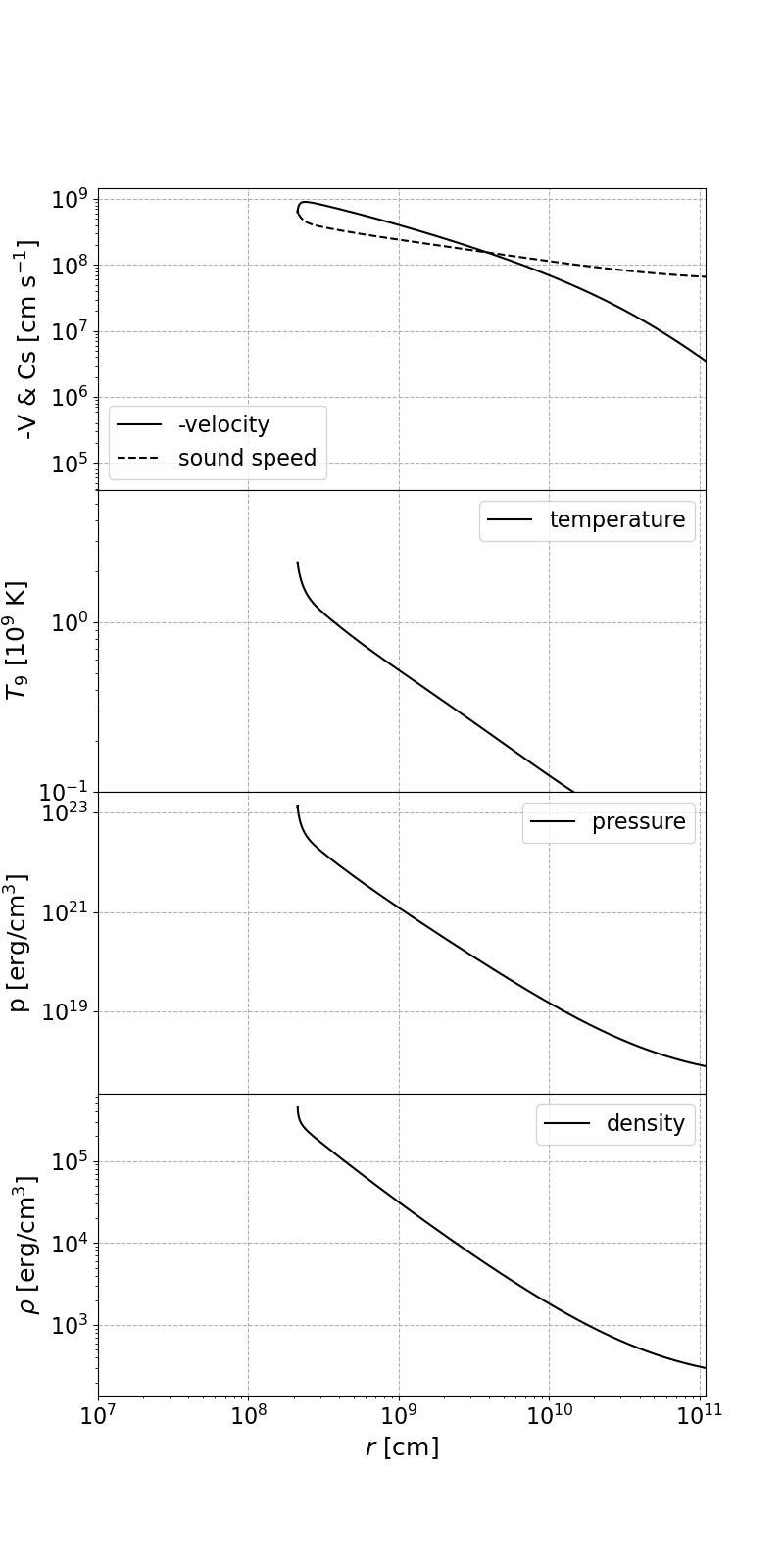}
 \end{center}
 \caption{The same as Figure \ref{f:pHe@sar} but with a larger accretion rate, $|\dot M|=1.6\times10^{32}$ g s$^{-1}$.}\label{f:pHe@lar}
\end{figure}

\begin{figure}
 \begin{center}
  \includegraphics[width=80 mm]{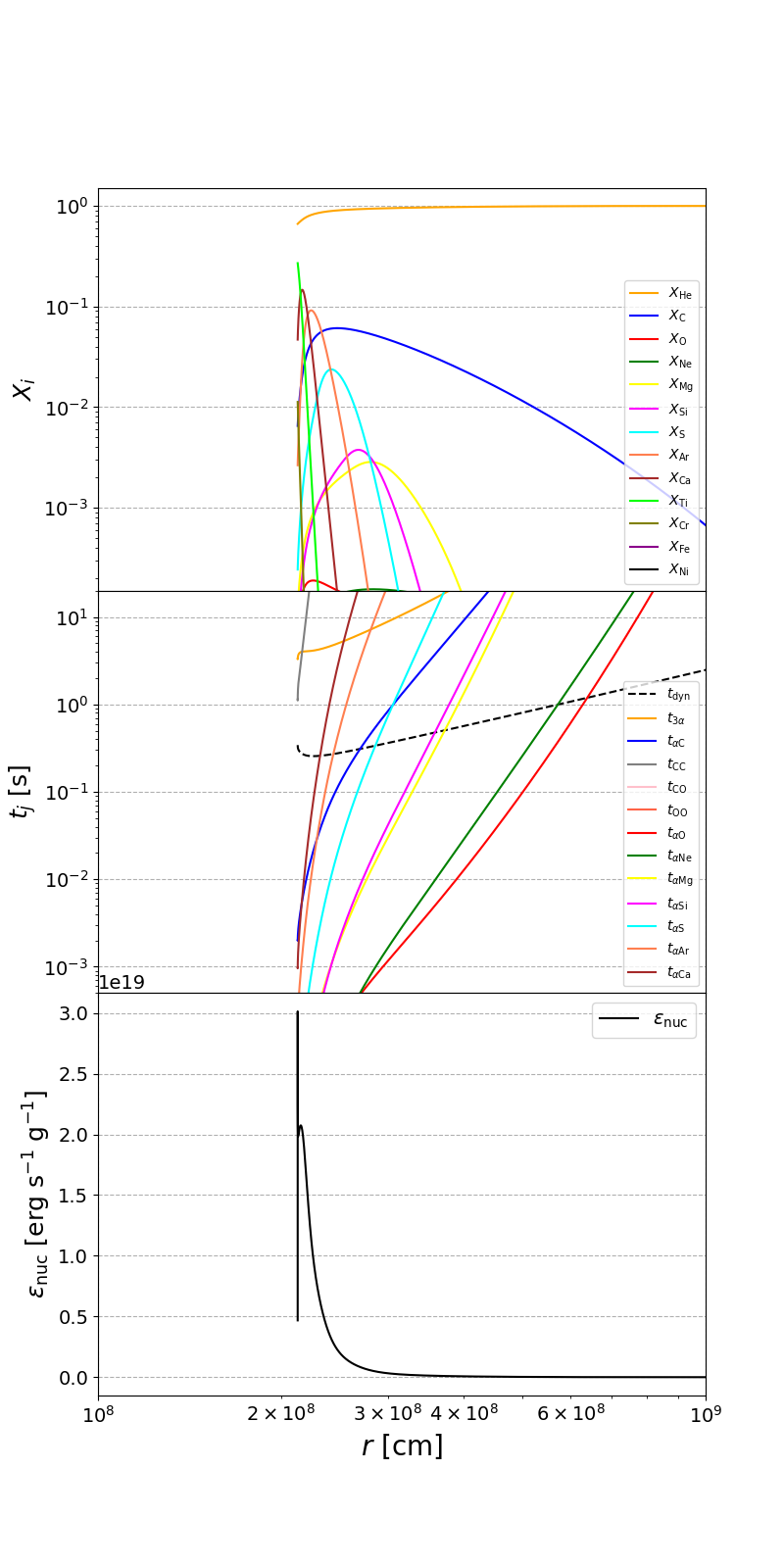}
 \end{center}
 \caption{The same as Figure \ref{f:pHe@sarN} but with a larger accretion rate, $|\dot M|=1.6\times10^{32}$ g s$^{-1}$.}\label{f:pHe@larN}
\end{figure}

If the accretion rate is increased by 60\% from the above model, i.e., $|\dot M|=1.6\times10^{32}$ g s$^{-1}$, then the flow decelerated by the energy from nuclear burning truncates where the absolute value of the velocity becomes equal to the sound speed. This point becomes singular in the sense that the derivatives in the governing equations diverge. Figures \ref{f:pHe@lar} and \ref{f:pHe@larN} show the distribution of physical quantities and the distributions of mass fraction of each element and the timescales of nuclear reactions in the model with this accretion rate, respectively. The flow truncates at the point where the timescale of C+C reactions decreases along the streamline down to the dynamical timescale. Note that the timescales of $\alpha$-capture reactions with heavier elements already become shorter than the dynamical timescale. Thus, the nuclear energy generated from these reactions most substantially affects the flow, even if the accreted matter is composed initially of pure helium. 

\begin{figure}
 \begin{center}
  \includegraphics[width=80 mm]{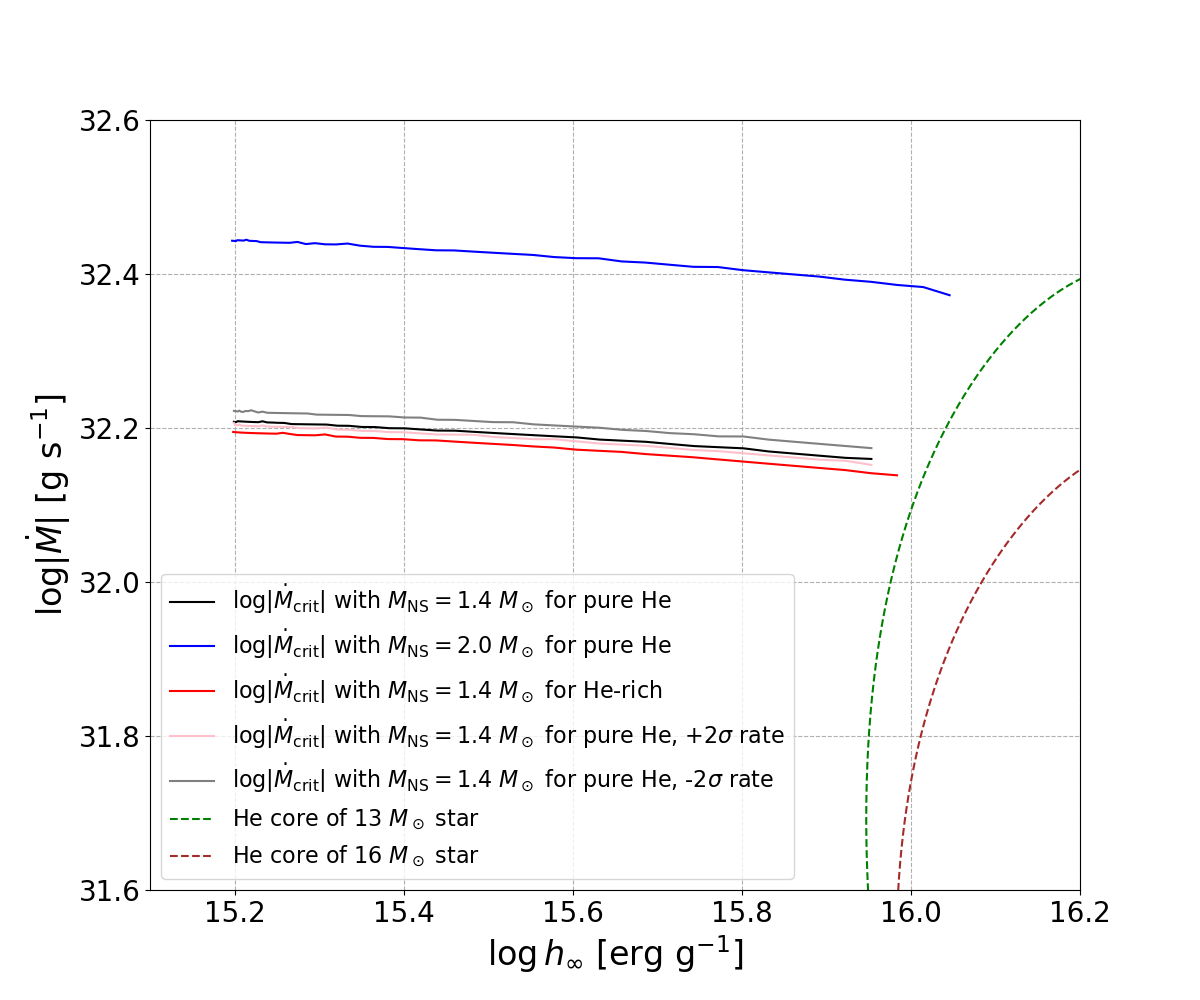}
 \end{center}
 \caption{The critical accretion rates as functions of the specific enthalpy in the ambient medium for $M=1.4\ M_\odot$ and 2 $M_\odot$. The pink (grey) line shows the critical accretion rates for the reaction rate of \atom{C}{}{12}($\alpha,\ \gamma$)\atom{O}{}{16} with +(-)2$\sigma$ deviation from the most probable values. The dashed lines show the expected accretion rates if the compact object with $M=1.4\ M_\odot$ resides in the He cores of three model stars with initial masses indicated in the legend. The accretion rates are estimated from Equation (\ref{eq:ar}).}\label{f:car_pHe}
\end{figure}

The critical accretion rate that separates types of the flow lies between $1.0\times10^{32}$ g s$^{-1}$ and $1.6\times10^{32}$ g s$^{-1}$ from these examples for this particular set of $h_\infty$ and $M$. We can obtain the critical accretion rate more precisely as a function of the specific enthalpy in the ambient medium and the mass of the compact object as was done in our previous work \citep{2022ApJ...933...29N}. The results for two masses ($M=1.4\ M_\odot$ and 2 $M_\odot$) of the compact object are shown in Figure \ref{f:car_pHe}. This figure also shows the expected accretion rates if the compact object resides in the helium cores of evolved massive stars with masses indicated in the figure. We use a simple formula for the Bondi accretion, i.e., 
\begin{equation}\label{eq:ar}
    \left|\dot M\right|=4\pi\left(GM\right)^2c_{\rm s\infty}^{-3/2}\rho_\infty,
\end{equation}
to calculate the expected accretion rate.
Here, the values of the sound speed $c_{\rm s\infty}$ and the mass density $\rho_\infty$ are taken from those in the helium cores of models obtained using the stellar evolution code MESA (r15140) \citep{Paxton11,Paxton13,Paxton15,Paxton18,Paxton19}. 

The assumption that nuclear burning affects the flow only in the supersonic region may not hold for the flow with a higher $h_\infty$ because the flow with  $h_\infty\gtsim10^{16}$ erg g$^{-1}$ truncates very close to the sonic point due to nuclear burning if the accretion rate is larger than the critical value. Our approach cannot be applied to such flows. Thus we do not draw the critical accretion rates at such high $h_\infty$ in this figure. Most of matter in the helium cores of the MESA models have too high specific enthalpies to directly apply our model because the accretion of this matter would result in significant nuclear burning in the subsonic region.

The flow truncates when the reaction timescale of C+C becomes comparable to the dynamical timescale as shown in Figure \ref{f:pHe@larN}, which was pointed out by \citet{2022ApJ...933...29N} for the accretion of CO matter. We will discuss the key reaction for the critical accretion rates of the flow with different compositions in section \ref{sec:conclusion}.


\subsection{Accretion of He-rich matter}\label{sec:He-rich}
As a next step, we investigate the accretion flow with more realistic compositions, i.e., He and some heavy elements. The abundances of the heavier elements roughly follow the solar abundances and the compositions are listed in Table \ref{tbl:ab_He-rich}.

\begin{table}
  \tbl{The assumed abundances of He-rich matter\footnotemark[$*$] }{%
  \begin{tabular}{ll|ll}
  \hline
    elements & mass fraction & elements & mass fraction \\ \hline\hline
    \atom{He}{}{4}&  0.9817 & \atom{Ar}{}{36}& 7.7$\times10^{-5}$ \\
    \atom{C}{}{12}& 0.0041 & \atom{Ca}{}{40}& 6.0$\times10^{-5}$ \\
    \atom{O}{}{16}& 0.0096 & \atom{Ti}{}{44}& 0.0 \\
    \atom{Ne}{}{20}& 0.0017 & \atom{Cr}{}{48}& 4.0$\times10^{-5}$ \\
    \atom{Mg}{}{24}& 0.00051 &\atom{Fe}{}{52}& 0.0 \\
    \atom{Si}{}{28}&  0.00065 &\atom{Ni}{}{56} & 0.0012 \\ 
    \atom{S}{}{32}& 0.00040 &  --& --\\ \hline
  \end{tabular}}\label{tbl:ab_He-rich}
  \begin{tabnote}
 \footnotemark[$*$] The abundances of heavy elements follow the solar abundances. The abundance of \atom{Fe}{}{56} is assigned to \atom{Ni}{}{56} since we do not consider the decay of \atom{Ni}{}{56}.
  \end{tabnote}
\end{table}

The inclusion of carbons in the accreted matter enhances the nuclear energy generation rate from C+C reactions and results in a few percent smaller critical accretion rates (Fig. \ref{f:car_pHe}). Though the essential features of the flow do not change from the pure He case, $\sim1.5$ \% increase in the C abundance at the peak leads to this reduction of the critical accretion rates.

\begin{figure}
 \begin{center}
  \includegraphics[width=80 mm]{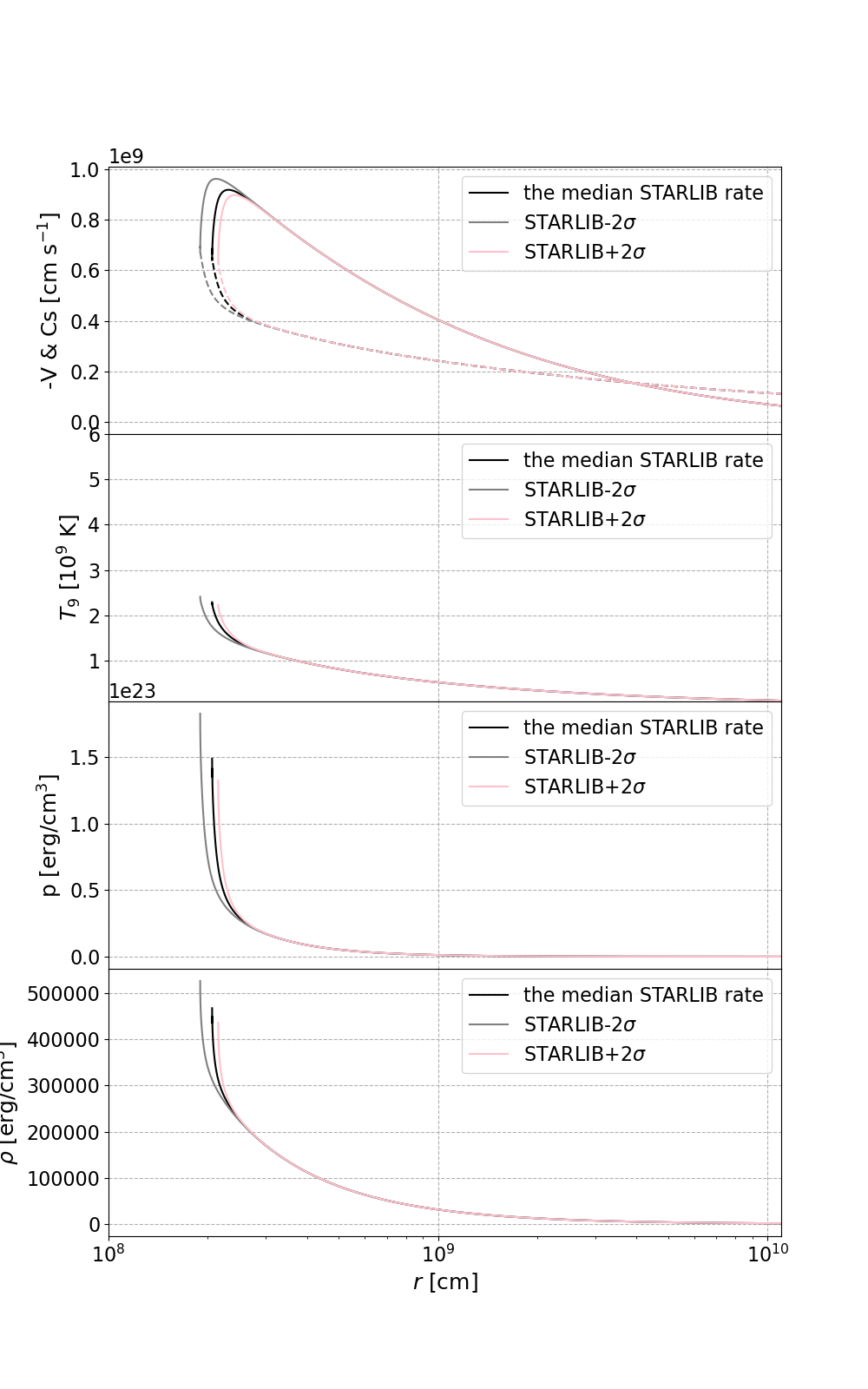}
 \end{center}
 \caption{Comparison of accretion flows with different reaction rates for \atom{C}{}{12}($\alpha,\gamma$)\atom{O}{}{16}. The parameters ($M$, $h_\infty$, and $\dot M$) are the same as those in Figure \ref{f:pHe@lar}.}\label{f:pHe@cmp}
\end{figure}

In our calculations, we use the rates of $\alpha$-capture reactions compiled by \citet{Cyburt_2010}. The rates such as \atom{C}{}{12}($\alpha,\ \gamma$)\atom{O}{}{16} might include significant uncertainties at high temperatures.   \citet{2024MNRAS.531.2786K} investigated the influence of uncertainty in the rate of  \atom{C}{}{12}($\alpha,\ \gamma$)\atom{O}{}{16} on pair-instability supernovae based on the approach taken by \citet{2010NuPhA.841....1L}. They assumed that each nuclear property follows a lognormal distribution with an uncertainty $\sigma_i$. Here, we take the same approach to investigate the influence of the uncertainty on the accretion flow. The uncertainty of this rate is derived from \citet{2002ApJ...567..643K} (see Appendix for details). Results of truncated flows with different $\alpha$-capture reaction rates are shown in Figure \ref{f:pHe@cmp}. A higher (lower) reaction rate truncates the flow at a larger (smaller) radius. The influence on the accretion of pure He (or He-rich matter) is not so drastic. Actually, a $+(-)2\sigma$ different rate changes the critical accretion rate by -1.7\% (+3.6\%) at most as seen from Figure \ref{f:car_pHe}. 

\subsection{Accretion of CO-rich matter}\label{sec:CO}
We have investigated the accretion flow of CO-rich matter onto a compact object in \citet{2022ApJ...933...29N}. Their treatment of nuclear reactions is quite simple. Only the C+C and O+O reactions and the corresponding energy generation rates were taken into account. In this paper, we check the validity of their treatment by solving the nuclear reaction network. The assumed abundances of the accreted matter are listed in Table \ref{tbl:ab_CO-rich}.

\begin{figure}
 \begin{center}
  \includegraphics[width=80 mm]{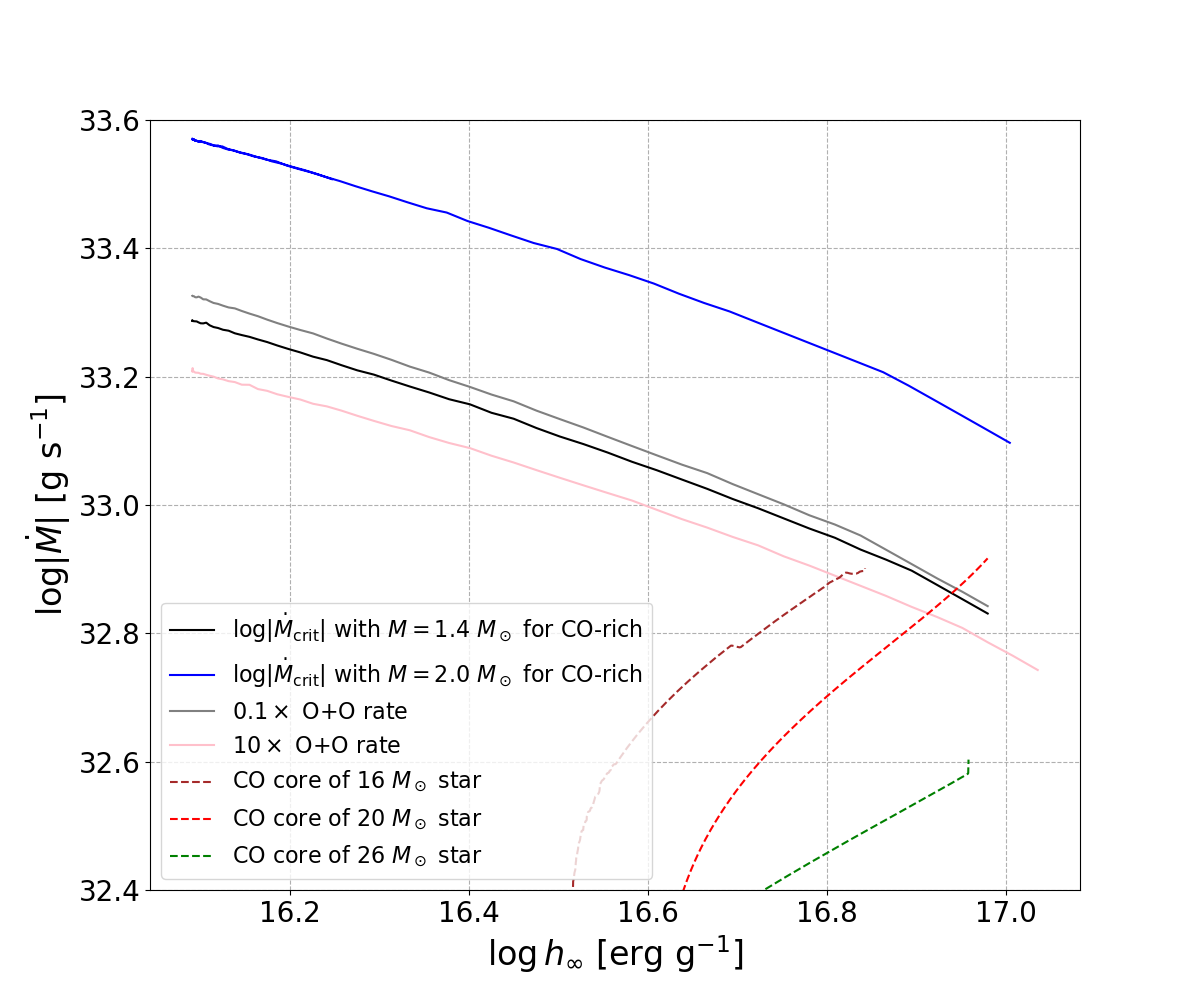}
 \end{center}
 \caption{The critical accretion rates as functions of the specific enthalpy in the ambient medium for $M=1.4\ M_\odot$ and 2.0 $M_\odot$ when the chemical composition of the ambient medium is given by those in Table \ref{tbl:ab_CO-rich}(the black solid line). The grey (pink) line shows the critical accretion rate with the reaction rate of O+O modified by a constant factor of 0.1 (10). The dashed lines show the expected accretion rates if the compact object resides in the CO cores of three model stars with initial masses indicated in the legend. The accretion rates are estimated from Equation (\ref{eq:ar}).}\label{f:car_CO}
\end{figure}

\begin{figure}
 \begin{center}
  \includegraphics[width=80 mm]{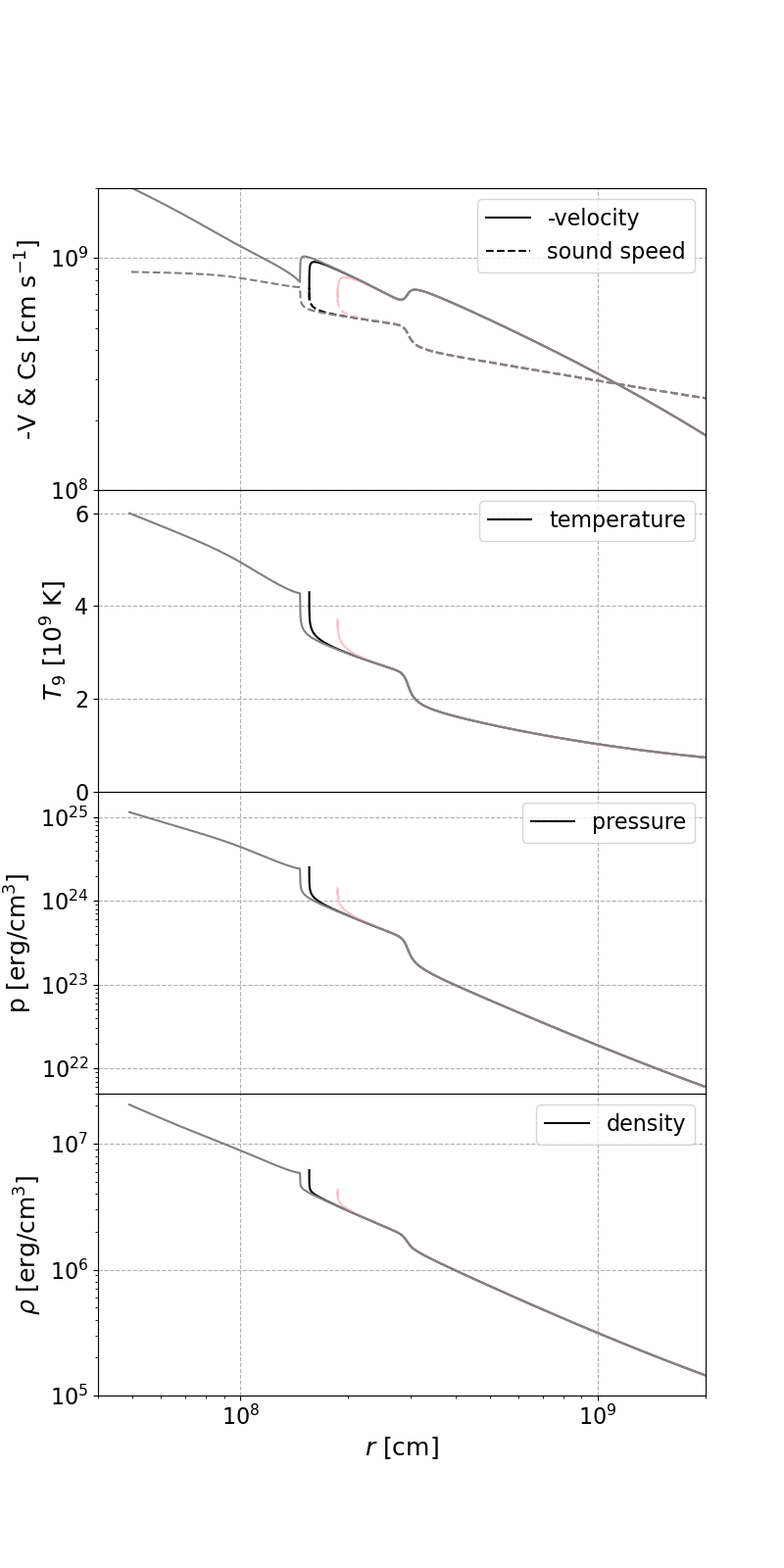}
 \end{center}
 \caption{Comparison of the accretion flows for the accretion of CO-rich matter with $M=1.4$ M$_\odot$, $|\dot M|=1.25\times10^{33}$g s$^{-1}$, and $\log h_\infty$ [erg g$^{-1}$]=16.60 with different O+O reaction rates. The grey (pink) solid line shows the critical accretion rate with the reaction rate of O+O multiplied by a constant factor of 0.1 (10) from the fiducial rate shown by the black solid line.}\label{f:CO@lar}
\end{figure}

\begin{figure}
 \begin{center}
  \includegraphics[width=80 mm]{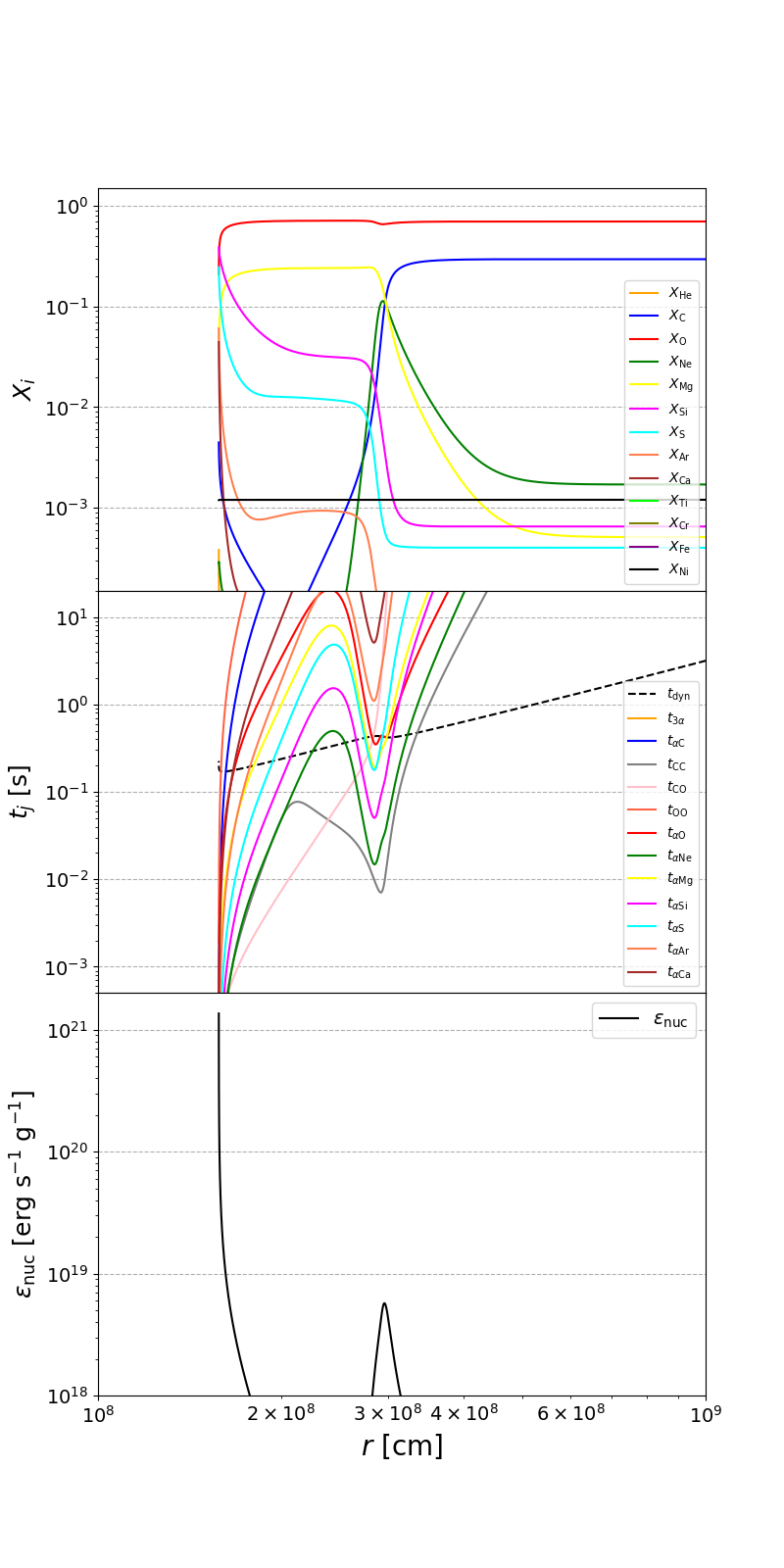}
 \end{center}
 \caption{The same as Figure \ref{f:pHe@sarN} but for the accretion of CO-rich matter with $M=1.4$ M$_\odot$, $|\dot M|=1.25\times10^{33}$g s$^{-1}$, and $\log h_\infty$ [erg g$^{-1}$]=16.60.}\label{f:CO@larN}
\end{figure}

\begin{figure}
 \begin{center}
  \includegraphics[width=80 mm]{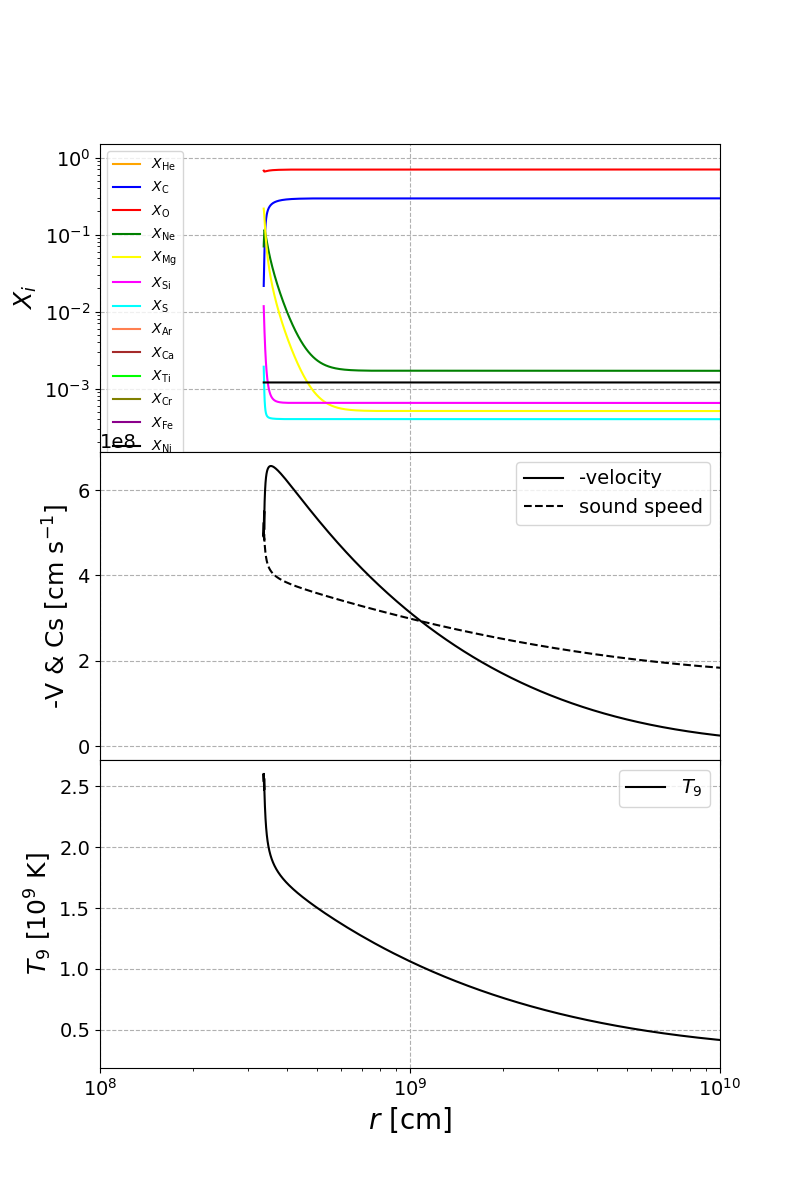}
 \end{center}
 \caption{The distributions of elements (top panel), the velocity and the sound speed (middle panel), and the temperature (bottom panel) as functions of radius for the accretion of CO-rich matter with $M=1.4$ M$_\odot$, $|\dot M|=1.7\times10^{33}$g s$^{-1}$, and $\log h_\infty$ [erg g$^{-1}$]=16.60.}\label{f:CO@slar}
\end{figure}

Figure \ref{f:car_CO} shows the critical accretion rates as functions of $h_\infty$. The values of the accretion rates become larger than those in Figure 5 of \citet{2022ApJ...933...29N} by a factor of $\sim2$. The empirical formula adopted in the previous work overestimates the energy generation rate due to O+O fusion reactions compared with that from the reaction network in the present calculations. We notice that the critical accretion rate is determined by the contribution from O+O reactions rather than C+C. This feature was already captured by calculations presented in Figure 3 of \citet{2022ApJ...933...29N} but was overlooked. In Figure \ref{f:car_CO}, the critical accretion rates for reaction rates of O+O enhanced (reduced) by a constant factor of 10 are also plotted as the pink (grey) line. Figures \ref{f:CO@lar} and \ref{f:CO@larN} illustrate that the flow truncates where the timescale of O+O reactions becomes comparable to the dynamical timescale. If the accretion rate is increased further, a flow truncated by C+C reactions appears(Fig. \ref{f:CO@slar}). 

At high specific enthalpies ($h_\infty\sim10^{17}$ erg g$^{-1}$), the flow truncates due to C+C reactions rather than O+O. This is why the reduced reaction rate of O+O  only slightly changes the critical accretion rate especially at high specific enthalpies as shown in Figure \ref{f:car_CO}. On the other hand, the enhanced O+O reaction rate gives lower critical accretion rates because the flow truncates due to O+O reactions even at high $h_\infty$. We do not plot the critical accretion rates at higher specific enthalpies ($h_\infty\gtsim10^{17}$ erg g$^{-1}$) because it seems that the flow is already affected by nuclear burning in the subsonic region. Thus our assumption of no nuclear burning in the subsonic region does not hold for such flows as is the case for the accretion of He-rich matter with high specific enthalpies.
\begin{table}
  \tbl{The assumed abundances of CO-rich matter\footnotemark[$*$] }{%
  \begin{tabular}{ll|ll}
  \hline
    elements & mass fraction & elements & mass fraction \\ \hline\hline
    \atom{He}{}{4}&  0.0 & \atom{Ar}{}{36}& 7.7$\times10^{-5}$ \\
    \atom{C}{}{12}& 0.295363 & \atom{Ca}{}{40}& 6.0$\times10^{-5}$ \\
    \atom{O}{}{16}& 0.7 & \atom{Ti}{}{44}& 0.0 \\
    \atom{Ne}{}{20}& 0.0017 & \atom{Cr}{}{48}& 4.0$\times10^{-5}$ \\
    \atom{Mg}{}{24}& 0.00051 &\atom{Fe}{}{52}& 0.0 \\
    \atom{Si}{}{28}&  0.00065 &\atom{Ni}{}{56} & 0.0012 \\ 
    \atom{S}{}{32}& 0.00040 &  --& --\\ \hline
  \end{tabular}}\label{tbl:ab_CO-rich}
  \begin{tabnote}
 \footnotemark[$*$] The abundances of elements heavier than oxygen follow the solar abundances. The abundance of \atom{Fe}{}{56} is assigned to \atom{Ni}{}{56} since we do not consider the decay of \atom{Ni}{}{56}.
  \end{tabnote}
\end{table}

\section{Critical accretion rates}
\begin{figure}
 \begin{center}
  \includegraphics[width=80 mm]{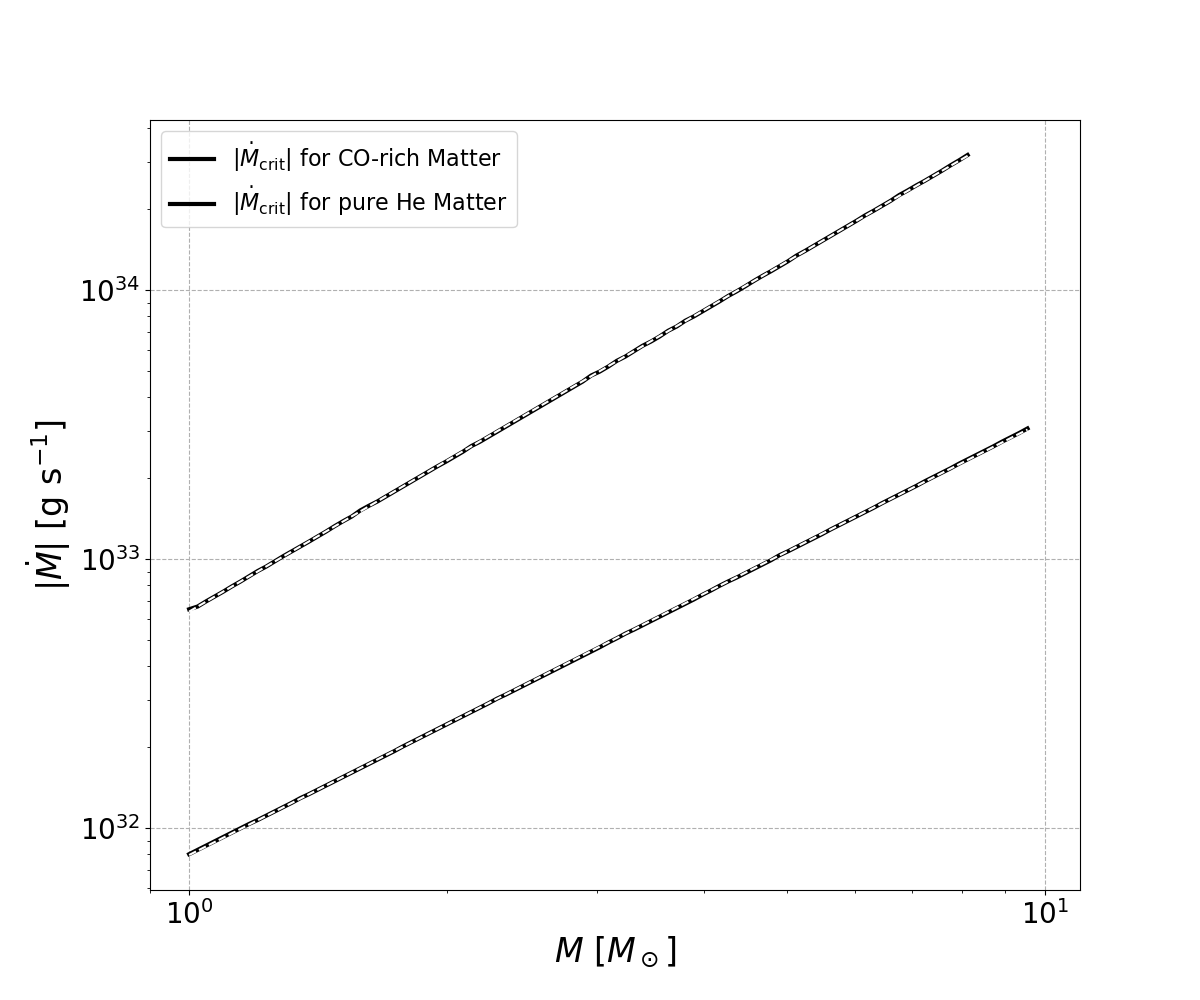}
 \end{center}
 \caption{The critical accretion rates for CO-rich matter ($h_\infty=3.658\times10^{16}$ erg g$^{-1}$) and pure He matter ($h_\infty=4.147\times10^{15}$ erg g$^{-1}$) as functions of the mass of the central object as indicated by the labels. Both of the rates can be fitted by power law functions of the mass. The exponents are 1.862 for CO-rich matter and 1.614 for pure He matter. The fitted functions are shown by white dashed lines.}\label{f:CAafM}
\end{figure}

We have summarized the critical accretion rates as functions of the mass $M$ of the central compact object in Figure \ref{f:CAafM} for different compositions of accreted matter.
It is clear from this figure that the dependence of the critical accretion rates on $M$ is different for different compositions. The critical accretion rates of He-rich matter are proportional to $M^{1.614}$ while those of CO matter are proportional to $M^{1.862}$. 

 By revisiting the analytical approach taken by \citet{2022ApJ...933...29N}, we derive an analytical expression for the critical accretion rate in this section. 

Suppose that the flow truncates at a radius $r=r_{\rm tr}$ where the dynamical timescale $t_{\rm dyn}$ can be defined as 
\begin{equation}
t_{\rm dyn}=\frac{r_{\rm tr}}{-v}\sim\sqrt{\frac{r_{\rm tr}^3}{GM}},
\end{equation}
since $v\sim-\sqrt{GM/r_{\rm tr}}$.
Our results suggest that the reaction timescale of C+C or O+O becomes comparable to the dynamical timescale at $r=r_{\rm tr}$:
\begin{equation}\label{eq:sigmav_c}
n_i(r_{\rm tr})<\sigma v>\sim\sqrt{\frac{GM}{r_{\rm tr}^3}},
\end{equation}
where $n_i(r_{\rm tr})$ denotes the number density of the relevant elements and $<\sigma v>$ is the rate of the key reaction, i.e., the rate of C+C reaction for He-
rich matter accretion and O+O for the CO-rich case. In terms of the mass density $\rho(r_{\rm tr})$, this can be rewritten as
\begin{equation}\label{eq:dtr}
\rho(r_{\rm tr})\sim\frac{A_i}{X_iN_{\rm A}<\sigma v>}\sqrt{\frac{GM}{r_{\rm tr}^3}},
\end{equation}
where $A_i$ and $X_i$ are the mass number and the mass fraction of the element $i$, respectively. Here the contribution from the initial thermal energy is ignored. The inclusion of the contribution from the initial thermal energy may reduce this density and the critical accretion rate. Though the dependence of the critical accretion rate on $h_\infty$ seen in Figures \ref{f:car_pHe} and \ref{f:car_CO} must be due to this contribution, the quantitative argument is beyond the scope of this orders-of-magnitude treatment. 

We can estimate the radius $r_{\rm tr}$ where the flow truncates as follows. If the accretion rate $|\dot M|$ is slightly greater than the critical value, the energy generated from the nuclear burning should become comparable to the gravitational energy near the truncation point. This is formulated as
\begin{equation}
    \epsilon_{\rm nuc}\sqrt{\frac{r_{\rm tr}^3}{GM}}=\frac{fGM}{r_{\rm tr}},
\end{equation}
where $f$ is a constant of the order of unity and the energy generation rate per unit mass $\epsilon_{\rm nuc}$ can be written as
\begin{equation}
    \epsilon_{\rm nuc}=n_i(r_{\rm tr})n_j(r_{\rm tr})<\sigma v>Q_{ij}/\rho(r_{\rm tr}).
\end{equation}
Here $Q_{ij}$ denotes the energy generated from a fusion reaction of elements $i$ and $j$. From these equations combined with equation (\ref{eq:sigmav_c}), one obtains 
\begin{eqnarray}
    r_{\rm tr}&=&\frac{fGMA_i}{X_iQ_{ij}N_{\rm A}}\\
    &=&\left\{\begin{array}{ll}
    5.4\times10^8f~{\rm cm}\left(\frac{A_i}{12}\right)\left(\frac{Q_{ij}}{14~{\rm MeV}}\right)^{-1}\left(\frac{X_i}{0.3}\right)^{-1}& \mbox{for C},\\ \nonumber
    2.7\times10^8f~{\rm cm}\left(\frac{A_i}{16}\right)\left(\frac{Q_{ij}}{16~{\rm MeV}}\right)^{-1}\left(\frac{X_i}{0.7}\right)^{-1}& \mbox{for O}.
    \end{array}\right.
\end{eqnarray}
If $f\sim0.6$, this formula reproduces the results for the accretion of CO-rich matter presented in section \ref{sec:CO}. Thus the truncation radius $r_{\rm tr}$ becomes roughly constant for given $X_i$. while this formula gives a too large radius for those of He-rich matter in sections \ref{sec:pHe} and \ref{sec:He-rich}. This is because the energy generated from $\alpha$-capture reactions is dominant over that from C+C in these flows. Actually, the timescales of $\alpha$-capture by various elements produced by carbon burning already become shorter than the dynamical timescale as shown in Figure \ref{f:pHe@larN}.

Substituting $\rho(r_{\rm tr})$ in Equation (\ref{eq:dtr}) into the formula of the accretion rate $\dot M=4\pi r_{\rm tr}^2\rho(r_{\rm tr})v$, one obtains the critical accretion rate
\begin{eqnarray}
\dot M_{\rm c}&\sim&-\frac{4\pi A_iGM}{X_i<\sigma v>N_{\rm A}}\\
&\sim&-10^{32}\ {\rm g}\ {\rm s}^{-1}\left(\frac{M}{1.4\ M_\odot}\right)\left(\frac{X_i<\sigma v>N_{\rm A}/A_i}{10^{-5}{\rm cm^3 s^{-1} mol^{-1}}}\right)^{-1}\nonumber.
\end{eqnarray}
Here the factor $X_i<\sigma v>/A_i$ should be evaluated for the carbon (oxygen) burning if we consider the accretion of He-rich (CO-rich) matter. These different key reactions result in different exponents of the critical accretion rates as power-law functions of $M$, though we cannot evaluate the exact values of the exponents from this approach.

In the accretion of He-rich matter, $X_i$ to be evaluated at the truncation point is the mass fraction of carbon produced by triple-$\alpha$ reactions in the upstream. The maximum value of $X_{\rm C}$ is determined by the balance between the rates of the triple-$\alpha$ reactions and the $\alpha$-capture reactions by C. Equating these rates, one obtains the maximum $X_{\rm C}$ in terms of the reaction timescales introduced in section \ref{sec:fof} as
\begin{equation}
    X_{\rm C,\ max}=3X_{\rm He}t_{\alpha \rm C}/t_{3\alpha}\sim3t_{\alpha \rm C}/t_{3\alpha}.
\end{equation}
Since the timescale of the $\alpha$-capture reaction is several tenths shorter than that of triple-$\alpha$ reactions at the peak of $X_{\rm C}$ as seen from the middle panels of Figures \ref{f:pHe@sarN} and \ref{f:pHe@larN}, the value of $X_{\rm C}$ becomes less than 0.1 at maximum. At the truncation point,  $X_{\rm C}$ decreases from the maximum because C+C reactions start to destruct C near the truncation point. The reactions responsible for truncating the flow are $\alpha$-capture reactions by heavy elements produced by C+C reactions, which become more abundant than C near the truncation point.

\section{Conclusions and discussion}\label{sec:conclusion}
We investigate the impacts of nuclear burning upon spherical stationary accretion flow onto a compact object especially when the accreted matter is He-rich. This situation is realized after a neutron star or black hole is engulfed by the He core of an evolved massive star. The development of time-domain astronomy will capture consequences of such rare events in near future. More generally and frequently occurring situations may be after the iron core of a massive star starts a gravitational collapse and the core accretes the He-rich (CO) envelope. We find the critical accretion rates as a function of the mass of the compact object and the specific enthalpy of the ambient matter as was found by \citet{2022ApJ...933...29N} for CO matter. The flow composed of He-rich matter with an accretion rate higher than the critical rate always truncates by the extra energy generated by $\alpha$-capture reactions. This is in contrast to the accretion of CO matter in which an accretion rate slightly higher than the critical rate truncates the flow due to the energy generated from O burning, while the flow with a still higher accretion rate is truncated by C burning. The understanding of the end results of these events is crucially dependent on our knowledge of the key nuclear reactions including \atom{O}{}{16}+\atom{O}{}{16} and \atom{C}{}{12}+\atom{C}{}{12}; In fact, the latter has been widely studied in the context of nuclear physics, such as the properties of low-energy resonances (e.g.,  \cite{2007RPPh...70.2149F}), nuclear potential models~\citep{2007PhRvC..76d5802G,2011PhRvC..84f4613E,2018PhRvC..97e5802D,2018PhRvC..98f4604C}, and recently uncertainties of nuclear effective interaction with antisymmetrized molecular dynamics~\citep{2021PhLB..82336790T,2024PhLB..84938434T}.


The numerical method taken in this paper assumes a flow in which the nuclear burning takes place only in the super-sonic region. Otherwise, the nuclear reaction network has to be solved for the entire flow, which makes it very difficult to find the transonic point. Investigation of such accretion flows of high-specific-enthalpy matter with explosive nuclear burning in the subsonic region is left for our future work. This might affect the yields of heavy elements from a core-collapse supernova and even the explosion itself though \citet{1966ApJ...143..626C} once denied such a possibility due to the dominance of the rarefaction wave. Nuclear burning in the subsonic region is likely to avoid the effects of the rarefaction wave and the current pre-supernova models suggest that the envelopes of massive stars have high specific enthalpies.

\section*{Acknowledgment}

We thank D. Kawashimo for providing us the data of ${}^{12}{\rm C}(\alpha,\gamma){}^{16}{\rm O}$ rate. This work is supported by JSPS KAKENHI Grant Numbers JP23K19056 (A.D.), JP22K03688, JP22K03671, and JP20H05639 (T.S.).

\appendix

\section*{Reaclib fitting formula of ${}^{12}{\rm C}(\alpha,\gamma){}^{16}{\rm O}$ rates}

\begin{figure}[h]
    \centering
\includegraphics[width=\linewidth]{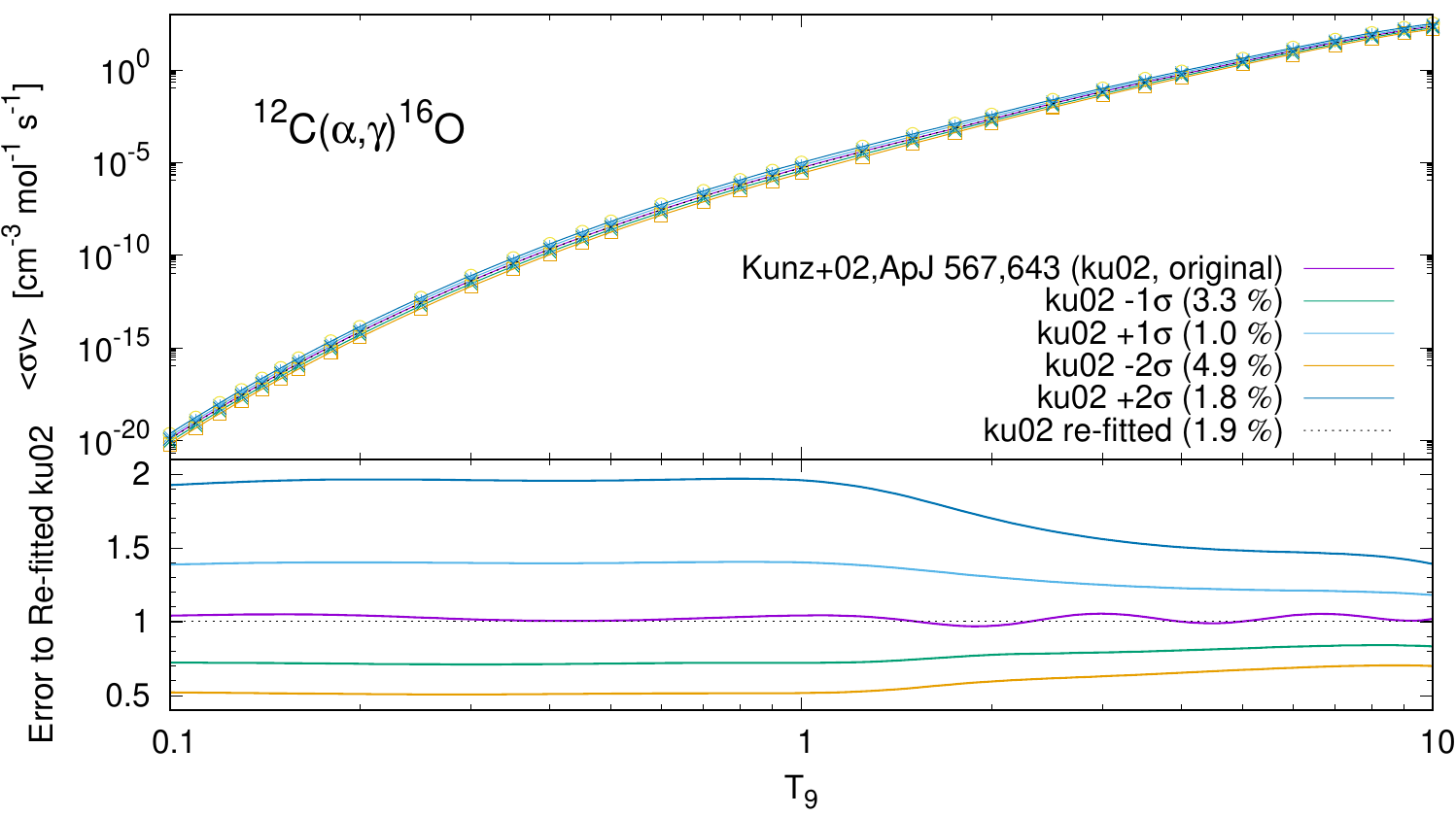}
    \caption{$T_9$-dependence of reaction rates of ${}^{12}{\rm C}(\alpha,\gamma){}^{16}{\rm O}$ (top) and relative error to \citet{2002ApJ...567..643K} we re-fitted (bottom). The fitting errors are expressed as percentages.}
    \label{fig:rate_unc}
\end{figure}

Data of ${}^{12}{\rm C}(\alpha,\gamma){}^{16}{\rm O}$ reaction rates in \citet{2024MNRAS.531.2786K} are based on the STARLIB library~\citep{2013ApJS..207...18S}. Meanwhile, we utilize all reaction rates as
the REACLIB fitting formulae~\citep{Cyburt_2010}, which are generally written in the form of 
\begin{eqnarray}
    \left<\sigma v\right>=\sum_{X=a,b,c,\cdots}\exp\left(X_0+\sum_{i=1}^5X_iT_9^{\frac{2i-5}{3}}+X_6\ln T_9\right),
\end{eqnarray}
as a function of the temperature $T_9$ in units of $10^9$ K. To take into account of uncertainties of the reaction rate of ${}^{12}{\rm C}(\alpha,\gamma){}^{16}{\rm O}$ based on \citet{2002ApJ...567..643K}, we fitted the original data of \citet{2024MNRAS.531.2786K} including some errors for the range of $0.04\le T_9\le10$ by the summation of two exponential functions with $X=a$ and $b$. The results are presented in Table \ref{tab:ku02}--\ref{tab:ku02+2s}. The reaction rates we re-fitted as a function of $T_9$ are shown in Figure~\ref{fig:rate_unc}.
All reaction rates utilized in this study can be successfully reproduced by the fitting formula within a few percent accuracy.

\begin{table}[t]
    \centering
    \caption{Fitting coefficients with \cite{2002ApJ...567..643K}. Note that the original rate implemented in the Jina Reaclib library is fitted with three exponential functions with $X=a$ and $b$.}
    \begin{tabular}{cccc}
    \hline\hline
         $a_0$     &     $a_1$  &        $a_2$   &       $a_3$ \\
           $a_4$    &      $a_5$   &       $a_6$ &   \\
 2.739963e+02 & -1.840100e+00 & 1.118288e+02 &-4.513673e+02 \\
 6.709498e+01 & -1.243748e+01 & 1.473981e+02 & \\
 \hline
           $b_0$     &     $b_1$  &        $b_2$   &       $b_3$ \\
           $b_4$    &      $b_5$   &       $b_6$ &   \\
 6.982260e+01 & -1.337762e+00 & 6.170204e+01 & -1.517089e+02\\
 9.014530e+00 & -5.192499e-01 & 7.292150e+01 &\\
 \hline \\
 \end{tabular}
 \label{tab:ku02}
    \centering
    \caption{Same as Table \ref{tab:ku02}, but with $-1\sigma$ errors from \cite{2002ApJ...567..643K} presented in \cite{2024MNRAS.531.2786K}.}
    \begin{tabular}{cccc}
    \hline\hline
           $a_0$     &     $a_1$  &        $a_2$   &       $a_3$ \\
           $a_4$    &      $a_5$   &       $a_6$ &   \\
 2.595149e+02 & -1.826437e+00 & 1.065764e+02 & -4.295560e+02\\
 6.466042e+01 & -1.250178e+01 & 1.406980e+02 &\\
 \hline
           $b_0$     &     $b_1$  &        $b_2$   &       $b_3$ \\
           $b_4$    &      $b_5$   &       $b_6$ &   \\
 6.879247e+01 & -1.344679e+00 & 6.072665e+01 & -1.499260e+02\\
 9.028506e+00 & -5.281048e-01 & 7.193656e+01 &\\
 \hline \\
 \end{tabular}
 \label{tab:ku02-1s}
    \centering
    \caption{Same as Table \ref{tab:ku02-1s}, but with $+1\sigma$ errors.}
    \begin{tabular}{cccc}
    \hline\hline
           $a_0$     &     $a_1$  &        $a_2$   &       $a_3$ \\
           $a_4$    &      $a_5$   &       $a_6$ &   \\
 2.681932e+02 & -1.820423e+00 & 1.082989e+02 & -4.408112e+02 \\
 6.613470e+01 & -1.240608e+01 & 1.436185e+02 & \\
 \hline
           $b_0$     &     $b_1$  &        $b_2$   &       $b_3$ \\
           $b_4$    &      $b_5$   &       $b_6$ &   \\
 7.042934e+01 & -1.333990e+00 & 6.024154e+01 & -1.505597e+02\\
 9.096262e+00 & -5.307871e-01 & 7.182692e+01 & \\
 \hline \\
\end{tabular}
\label{tab:ku02+1s}
    \centering
    \caption{Same as Table \ref{tab:ku02-1s}, but with $-2\sigma$ errors.}
    \begin{tabular}{cccc}
    \hline\hline
           $a_0$     &     $a_1$  &        $a_2$   &       $a_3$ \\
           $a_4$    &      $a_5$   &       $a_6$ &   \\
 2.606781e+02 & -1.826475e+00 & 1.076555e+02 & -4.322484e+02\\
 6.477957e+01 & -1.250674e+01 & 1.418088e+02 \\
 \hline
           $b_0$     &     $b_1$  &        $b_2$   &       $b_3$ \\
           $b_4$    &      $b_5$   &       $b_6$ &   \\
 6.838210e+01 & -1.339681e+00 & 6.121560e+01 & -1.503222e+02\\
 9.009086e+00 & -5.260212e-01 & 7.235348e+01 & \\
 \hline
 \end{tabular}
 \label{tab:ku02-2s}
    \centering
        \caption{Same as Table \ref{tab:ku02-1s}, but with $+2\sigma$ errors.}
    \begin{tabular}{cccc}
    \hline\hline
           $a_0$     &     $a_1$  &        $a_2$   &       $a_3$ \\
           $a_4$    &      $a_5$   &       $a_6$ &   \\
 2.654369e+02 & -1.784899e+00 & 1.054906e+02 & -4.344172e+02\\
 6.548171e+01 & -1.230900e+01 & 1.410761e+02 & \\
 \hline
           $b_0$     &     $b_1$  &        $b_2$   &       $b_3$ \\
           $b_4$    &      $b_5$   &       $b_6$ &   \\
 7.218321e+01 & -1.296553e+00 & 5.802949e+01 & -1.499280e+02\\
 9.270169e+00 & -5.499970e-01 & 7.062160e+01 &\\
 \hline
\end{tabular}
    \label{tab:ku02+2s}
\end{table}

\newpage

\bibliographystyle{apj}
\bibliography{references}{}

\end{document}